\documentclass[epj-spec]{svjour}
\usepackage{graphics}
\usepackage{bm}
\usepackage{amsmath}
\usepackage{epsfig}
\begin{document}
\title{Chiral symmetry, Confinement and Nuclear Matter properties }
\subtitle{}
\author{G. Chanfray\inst{1}\fnmsep\thanks{\email{g.chanfray@ipnl.in2p3.fr}} \and M. Ericson \inst{1,2} \and  E. Massot\inst{1} }
\institute{IPN Lyon, Universit\'e de Lyon/Univ. Lyon 1, IN2P3/CNRS \and Theory Division,CERN}
\abstract{
We discuss the possible influence of fundamental QCD properties such as spontaneous chiral symmetry breaking and nucleon substructure
on nuclear matter properties. We propose a chiral version of the relativistic $\sigma-\omega$ model in which the attractive background scalar field is associated with the chiral invariant field governing the radial fluctuations of the quark condensate. Nuclear matter 
stability is ensured once the scalar response of the nucleon depending on the quark confinement mechanism is properly incorporated. The needed parameters are estimated from lattice results and a satisfactory description of bulk properties follows, the only really free parameter being the
$\omega NN$ coupling constant. Pion loops can be also incorporated to obtain in a consistent way the finite density chiral susceptibilities.
A good description of the  asymmetry energy is obtained once the  full rho meson exchange and Fock terms are included. 
} 
\maketitle
\section{Introduction}
\label{intro}
A fundamental question of present day theoretical nuclear physics is the relationship between non perturbative QCD and the very rich structure of the nuclear many-body problem. A first trial beyond the standard non relativistic treatment based on effective forces (Skyrme or Gogny forces) is the relativistic mean field approach initiated by Walecka \cite{SW86}. The next step should be  to incorporate in this framework the effect associated 
with the most prominent properties of low energy QCD, {\it i.e.}, the spontaneous breaking of chiral symmetry and the effect of hadron substructure and color confinement. In a bottom-to-top attitude this certainly constitutes a necessary basis to study matter under extreme density and temperature conditions.
\section{Chiral symmetry and confinement}
\label{sec:1}

{\bf\it Basics on chiral symmetry}. The $SU(2)_L\otimes SU(2)_R$ chiral symmetry of the QCD Lagrangian is certainly 
a crucial key for the understanding of many phenomena in low energy hadron physics.
This symmetry originates from the fact that the QCD Lagrangian is almost invariant 
under the separate flavor $SU(2)$ transformations  of left-handed $q_L=(u_L,d_L)$
and  right-handed $q_R=(u_R,d_R)$ light quark fields $u$ and $d$.
The explicit violation of chiral symmetry is governed by the quark mass 
$m_q=(m_u+m_d)/2\le$ 10 MeV which is much smaller than typical hadron masses 
of order $1$ GeV, indicating that the symmetry is excellent. It is however well established 
that the QCD vacuum does not possess the symmetry of the Lagrangian {\it i.e.}, chiral
symmetry is spontaneously broken (SCSB) as it is evidenced by a set of remarkable properties. 
The first one is the building-up of a
chiral quark condensate~: $\langle \bar q q \rangle =\langle \bar u u +\bar d d \rangle/ 2$
which mixes, in the broken vacuum, left-handed and right-handed
quarks ($\langle \bar q q \rangle =\langle \bar q_L q_R + \bar q_R q_L\rangle/2$).
Another order parameter at the hadronic scale is the pion decay constant
$f_\pi=94$ MeV which is related to the quark condensate by the
Gell-Mann-Oakes-Renner (GOR) relation~: $-2 m_q \langle\bar q q\rangle_{vac}=m^2_\pi f^2_\pi$ 
valid to leading order in the current quark mass. It leads to 
a large negative value $\langle\bar q q\rangle_{vac}\simeq$ -(240 MeV)$^3$ 
indicating a strong dynamical breaking of chiral symmetry. The second feature is the
appearance of soft Goldstone bosons to be identified with the almost massless pions.
Finally the chiral asymmetry of the vacuum associated with the condensation of
quark-antiquark pairs is directly visible at the level of the the hadronic
spectrum~: there is no degeneracy between possible chiral partners such as 
$\rho(770)-a_1(1260)$ or $\pi(140)-\sigma(400-1200)$. 

\bigskip\noindent
{\it An illustration: the Nambu-Jona-Lasinio (NJL)  model}. Although chiral symmetry breaking and restoration can be studied on the lattice, the 
underlying physical mechanisms at the QCD level are not yet fully understood.  It is thus useful to study this problem with models such as the very popular NJL model. In such a model the very complicated highly non perturbative multi-gluons exchanges between quarks are simulated by a very simple chiral invariant contact interaction with strength parameter $G_1$.  In its simplest form the Lagrangian is~: 
\begin{equation}
{\cal L}_{NJL}=i\bar\psi\gamma^\mu\partial_\mu\psi \,-\,m\,\bar\psi \psi\,+\,
{G_1\over 2}\left[(\bar\psi \psi)^2\,+\,
\left(i\bar\psi\gamma^5 \vec\tau \psi\right)^2\right].
\end{equation}
where $\psi=\left(\psi_u, \psi_d\right)$ is an isodoublet of quark fields. The model also incorporates the fact that only low momentum quarks strongly interact in QCD. This is achieved by including a cutoff $\Lambda$ of the order of $1$ GeV in the momentum expansion of the quark fields. It is very easy to show that if the strength $G_1$ is sufficiently large, the ground state wave function is of BCS type~, $|\phi(M)\rangle= C\,exp\left(-\sum_{s,\, p<\Lambda}\,
\gamma_{ps}\,b^\dagger_{{\bf p} s}\,d^\dagger_{-{\bf p}-s}\right) |\phi_0\rangle$, and made of interacting quark-antiquark pairs (created by the $b^\dagger$ and $d^\dagger$ operators acting on the bare perturbative vacuum $|\phi_0\rangle$), hence building up  a quark condensate. As well-known such a BCS ground state is the vacuum of quasi-particles, the constituant quarks, with mass $M_Q$ solution of a gap equation~:
\begin{equation}
M_Q=m\,-\,2 G_1\,\langle\langle\bar q q\rangle\rangle
= m\,+\,4\,N_c\,G_1\,\int_{p<\Lambda}\,
{d{\bf p}\over (2\pi)^3}\,{M_Q\over E_p},\qquad E_p=\sqrt{M^{2}_{Q}+p^2}.\label{GAP}	
\end{equation}
It is thus clear that the existence of such a constituant mass of the order of $350$ MeV is intimately related to the quark condensate
$\langle\langle\bar q q\rangle\rangle=\langle\langle\bar\psi_u \psi_u + \bar\psi_d \psi_d\rangle\rangle/2$ which plays the role of an order parameter associated with the broken vacuum.

\smallskip\noindent
The mesons can be obtained using standard RPA exactly as in nuclear physics,  as collective quark-antiquark excitations. This concerns in first place the pion (generated in the $\bar\psi\gamma^5 \vec\tau \psi$ channel) which appears as a Goldstone boson and its chiral partner the scalar-isoscalar sigma meson ($\bar\psi \psi$ channel) which comes out with a mass $m_\sigma\simeq2\,M_Q$. This model insisting on chiral symmetry, once  taken in a more sophisticated form, yields a rather good phenomenology, especially in the meson sector, but has an important weakness since it does not incorporate color confinement \cite{R97}.

\bigskip\noindent
{\it Interplay between chiral symmetry and confinement}. When hadronic matter is heated and compressed, initially confined quarks and 
gluons start to percolate between the hadrons to be finally liberated. This picture is supported by lattice simulations showing that strongly interacting matter exhibits a sudden change in thermodynamic quantities (constituting a true phase transition or a rapid cross-over) within a narrow temperature window around $T_c=170$ MeV. This transition is accompanied by a sharp dropping of the chiral condensate indicating chiral symmetry restoration. However why these two transitions seem to occur simultaneously is still a major  open question. This not fully understood duality 
between chiral symmetry and confinement is also present at the level of the nucleon structure and consequently in nuclear physics. Some of the pictures or models for  nucleon structure put the emphasis on color confinement. This is particularly true for the MIT bag model where  quarks with current quark mass, $m_q\simeq 5-10$ MeV,  move freely in a cavity (bubble of perturbative vaccuum)  generated by the confinement dynamics. A totally different view is provided by the   constituant quark picture where the nucleon is made of three "big" quarks ({\it i.e.}, of the NJL type) getting their mass from the chiral condensate of the non perturbative QCD vacuum. A plausible hybrid picture  mixing both aspects seems to be supported by lattice calculations \cite{B05}. The nucleon looks like a $Y$ shaped color string with three constituant quarks attached at the end-points of the $Y$. The introduction of  a coupling between quarks and mesons allows to build the pion cloud which plays an extremely important role to account for the mass and the chiral properties of the nucleon (such as the sigma term or  chiral susceptibilities defined below). Finally possible  configurations where a string develops between one quark and one diquark in the color antitriplet channel, might also be present in the nucleon.

\section{Chiral effective theory: nucleon structure and nuclear physics}
\label{sec:2} 
{\it How to build an effective chiral theory? NJL model as an example}. For applications in hadronic or nuclear physics, one usually tries to formulate low energy QCD directly in terms of physical degrees of freedom, namely the hadrons. Mathematically one has to integrate out quarks and gluons in favor of hadrons. This is of course a formidable task that we will illustrate with the example of the NJL model. The appropriate technics is the path integral formalism where the partition function (vacuum persistence amplitude)  is expressed as a functionnal integral over the quark fields:
$Z=\int\,D\bar{\psi}D\psi \,exp\left(\,i\,d^4 x{\cal L}_{NJL} \right)$. The idea is to make a change of variable in the functionnal integral 
$\sigma=g\,\bar\psi \psi\,,{\vec\pi}=g\,i\bar\psi\gamma^5 \vec\tau \psi$ ($g$ is a coupling constant to be fixed) and to integrate out the quarks in the Dirac sea. One thus obtains a very complicated highly non local effective Lagrangian (the fermion determinant) depending on the mesonic fields, $\sigma, \vec\pi,...$. An approximative form can be obtained using a derivative or a loop expansion. To one loop, one gets: 
${\cal L}_{eff}=(1/2)\left(\partial^\mu\sigma\partial_\mu\sigma + \partial^\mu\vec\pi\cdot\partial_\mu\vec\pi\right)-U(\sigma, \vec{\pi})$. 
The effective potential $U(\sigma, \vec{\pi})$ has the typical mexican hat shape associated with a broken (chiral) symmetry (see  fig. 1).
The degenerate vacuum corresponds to the chiral circle, {\it i.e.}, the bottom of the mexican hat (fig.1). The pseudoscalar field (pion) represents the phase fluctuation of the condensate and it is massless since it does not cost energy to move on the bottom. The massive scalar mode (sigma) has a mass related to the curvature of the potential and represents the amplitude fluctuation of the condensate. The vacuum point with isospin zero is the point where $\pi=0$ and where $\sigma$ has an expectation value  identified with the pion decay constant $f_\pi$. 
                
\begin{figure}
\noindent
 \begin{minipage}[b]{.30\linewidth}
    \centering\epsfig{figure=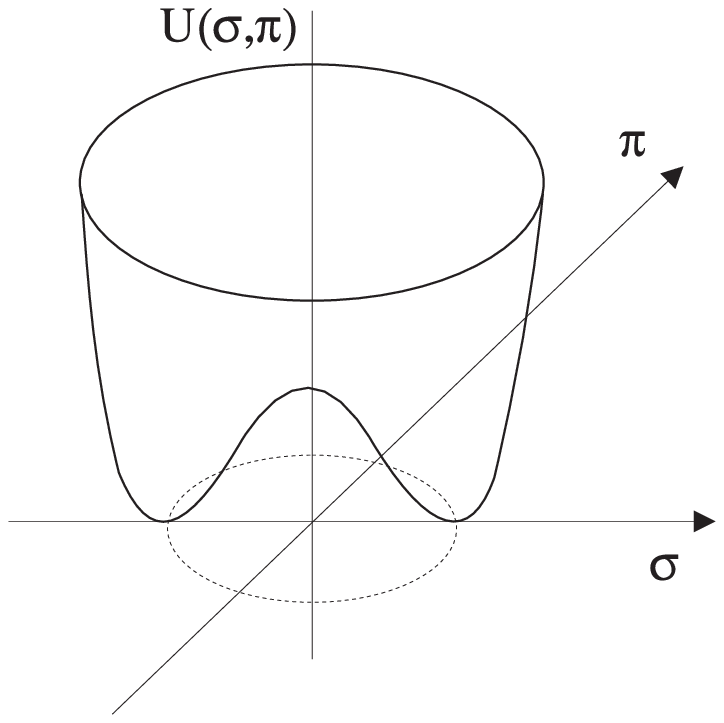,width=\linewidth}
  \end{minipage}\hfill
  \begin{minipage}[b]{.40\linewidth}. 
  \centering\epsfig{figure=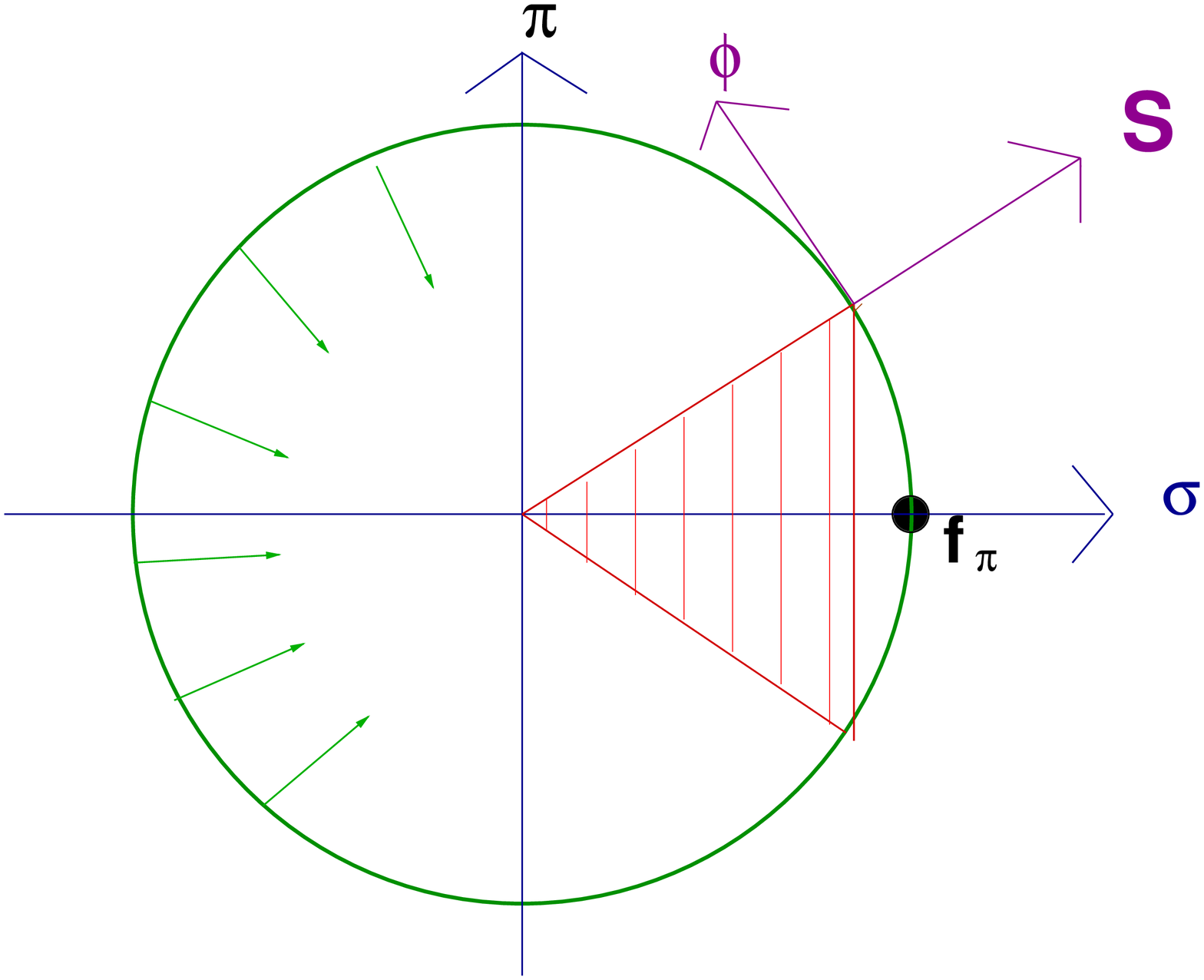,width=\linewidth}
  \end{minipage}
\caption{\it Left panel: Mexican hat effective potential. Right panel: Chiral circle corresponding to the bottom of the mexican hat potential } 
\end{figure}

\smallskip\noindent
{\it Effective chiral theory and the background scalar field in nuclear matter}. More generally we take the point of view that the effective theory has to be  formulated first in term of the fields associated with the fluctuations of the condensate parametrized in a matrix form $W=\sigma + i\vec\tau\cdot\vec\pi$.  The sigma and the pion are promoted to the rank of effective degrees of freedom. An alternative and very convenient formulation of the resulting sigma model is obtained by going from cartesian to polar coordinates (right panel of fig.1), {\it i.e.}, going from a linear to a non linear representation,  according to~:
\begin{equation}
W=\sigma\, + \,i\vec\tau\cdot\vec\pi=S\,U=(f_\pi\,+\,s)\,exp
\left({i\vec\tau\cdot\vec\phi\over f_\pi}\right).
\end{equation}
The new pion field $\vec\phi$ corresponds to an orthoradial soft mode which is automatically massless (in the absence of explicit CSB) since it is associated with rotations on the chiral circle without cost of energy. The new sigma meson field $S$, which is a chiral invariant,  
describes a radial mode associated with the fluctuations of the ``chiral radius'' $f_\pi$. It can be associated with the ordinary sigma meson 
which gets a very large width from its strong decay into two pions. Since it has derivative couplings, it decouples 
 from low energy pions whose dynamics is described by chiral perturbation theory. With increasing density, its fluctuations $s=S-f_\pi$ are associated with the shrinking of the chiral circle and it governs the evolution of the nucleon mass. In a recent paper  \cite{CEG02} we have proposed to identify this chiral invariant $s$ field with the sigma meson of nuclear physics and relativistic theories of the Walecka type, or, said differently, with the background attractive scalar field at the origin of the nuclear binding. This also gives a plausible solution to the long-termed problem of the chiral status of Walecka theories.

\bigskip\noindent
{\it Tests of the effective theory with a chiral invariant scalar field}. Once the appropriate couplings of the chiral fields to the baryons are introduced one can build an effective lagrangian to describe nuclear matter. Vector mesons ($\omega$ and $\rho$) must be also included to get the needed short range repulsion and asymmetry properties. At the Hartree level, the pion and the rho do not contribute for symmetric nuclear matter whose energy density written as a function of the order parameter $\bar s=\langle s \rangle$ is~:
\begin{equation}
{E_0\over V}=\varepsilon_0=\int\,{4\,d^3 p\over (2\pi)^3} \,\Theta(p_F - p)\,E^*_p(\bar s)
\,+\,V(\bar s)\,+\,{g^2_\omega\over 2\, m_\omega^2}\,\rho^2.\label{HARTREE}
\end{equation}
 $E^*_p(\bar s)=\sqrt{p^2\,+\,M^{*2}_N(\bar s)}$ is the energy of an 
effective nucleon with the effective mass $M^*_N(\bar s)=M_N +g_S\,\bar s$. $g_S$ is the scalar coupling constant of the model; in the pure linear sigma model it is $g_S=M_N/f_\pi$. The effective potential $U(\sigma, \vec{\pi})$ when reexpressed in term of the new representation has the form~:
$$V(s)=\frac{1}{2} m^{2}_{\sigma}\,\left(s^2\,+\,\frac{1}{2}\,\frac{s^3}{f_\pi}\,+...\right).$$
$\bar s$ is obtained by minimization of the energy density and is given at low density by~: $\bar s\approx-(g_S/ m_\sigma^2)\,\rho_S$. Its negative value is at the origin of the binding but the presence of the $s^3$ term (tadpole) has very important consequences. This tadpole is at the origin of the chiral dropping of the sigma mass $\Delta m^{*}_{\sigma} \simeq  -(3\,g_S /2  f_{\pi})\rho_S$ (a $\simeq 30\%$ effect at $\rho_0$), and generates an attractive three-body force which makes nuclear matter collapse and destroys the Walecka saturation mechanism. Hence the chiral theory does not pass the nuclear matter stability test. 

\smallskip\noindent
This failure, which is in fact a long-standing problem \cite{KM74,BT01}, is maybe not so surprising since  the theory, as it is, also fails to describe some nucleon structure aspects as discussed below. The nucleon mass, as well as other intrinsic properties of the nucleon (sigma term, chiral susceptibilities),
are QCD quantities which are in principle obtainable from lattice simulations. The problem is that lattice calculations of this kind are not feasible for quark masses smaller than $50$ MeV, or equivalently pion mass smaller than $400$ MeV, using the GOR relation. Hence one needs a technics to extrapolate the lattice data to the physical region. The difficulty of the extrapolation is linked to the non analytical behaviour of the nucleon mass as a function of $m_q$ (or equivalently $m^2_\pi$) which comes from the pion cloud contribution. The idea of Thomas {\it et al} \cite{TGLY04} was to separate the pion cloud self-energy, $\Sigma_{\pi}(m_{\pi}, \Lambda)$, from the rest of the nucleon mass and to calculate it in a chiral model   with  one adjustable cutoff parameter $\Lambda$. They expanded the remaining part  in terms of $m^2_{\pi}$  as follows~:
\begin {equation}
M_N(m^{2}_{\pi}) = 
a_{0}\,+\,a_{2}\,m^{2}_{\pi}\, +\,a_{4}\,m^{4}_{\pi}\,+\,\Sigma_{\pi}(m_{\pi}, \Lambda)\,.
\end{equation}
The best fit value of the parameter $a_{4}$   shows little sensitivity to the shape of the form 
factor, with a value 
$a_4 \simeq- 0.5\, \mathrm{GeV}^{-3} $ while $a_2 \simeq 1.5\,\mathrm{GeV}^{-1}$ (see ref. \cite{TGLY04}). The small value 
of $a_4$ reflects the fact that the non pionic contribution to the nucleon mass is almost linear in $m^2_{\pi}$ ({\it i.e.}, in $m_q$).
Taking successive derivatives of $M_N$ with respect to $m^2_{\pi}$ ({\it i.e.}, to $m_q$), it is possible to obtain some fundamental chiral properties of the nucleon,  namely the pion-nucleon sigma term and the scalar susceptibility of the nucleon. The non pionic pieces of these  quantities are given by~:
\begin{equation}
 \sigma_N^{non-pion} =m_q\, \frac{\partial M_N^{non-pion}}{\partial m_q}\simeq m^2_{\pi} {\partial M\over \partial m^2_{\pi }}
=a_2 \,m^2_{\pi}\, + \,2\,a_4 \, m^4_{\pi}\simeq 29\, \mathrm{MeV}\,.
\end{equation}
\begin{equation}
 \chi_{NS}^{non- pion}=\frac{\partial\left(\sigma_N^{non-pion}/2\,m_q\right)}{\partial m_q}\simeq 2{\langle\bar q q\rangle_{vac}^2 \over f^4_{\pi} }
 {\partial~~\over \partial m^2_{\pi }}\left({\sigma_N^{non-pion} \over m^2_{\pi }}\right)=  
 {\langle\bar q q\rangle_{vac}^2 \over f^4_{\pi} }\,4\,a_{4}\,.
\end{equation}
In the above equations the first equalities correspond to the definitions, the second equalities make use of the GOR relation and the last ones 
come from the lattice QCD analysis. With 
typical  cutoff used in this analysis, $\Lambda\simeq 1$ GeV, which  yields $\sigma_N^{(\pi)}\simeq20$ MeV,  the 
total value of the sigma term is $\sigma_N\simeq 50 $ MeV, a quite satisfactory result in view of the most recent analysis.
It is interesting to compare what comes out from the lattice approach with our chiral effective model. At this stage the only non pionic contribution 
to the nucleon mass comes from the scalar field, or more microscopically the nucleon mass entirely comes from the chiral condensate 
since the nucleon is just made of three constituant quarks with mass $M_Q=g\langle S \rangle_{vac}=g f_\pi\simeq 350$ MeV. Hence the results for the non pionic sigma term and scalar susceptibility are identical to those of the linear sigma model~:
\begin{equation}
\sigma_N^{(\sigma)}=f_\pi\,g_S\,\frac{m^2_\pi}{m^2_\sigma},\qquad
\chi^{(\sigma)}_{NS}=-2\,\frac{\left\langle \bar q q\right\rangle^2_{vac}}{f_\pi^3}\,
\frac{3\,g_S}{m^4_\sigma}	
\end{equation}
The identification of   $\sigma_N^{non-pion}$ with $\sigma_N^{(\sigma)}$ of our model
fixes the sigma mass to a value  $m_\sigma=800$ MeV, close to the one $\simeq750$ MeV that we have used 
in a previous article \cite{CE05}. As it is the ratio $g_S/{m^2_\sigma}$ which is thus determined this 
value of $m_\sigma$ is associated with the coupling constant of the linear sigma model 
$g_S=M_N/f_\pi=10$. Lowering $g_S$ reduces $m_\sigma$. Similarly, the comparison of $\chi^{(\sigma)}_{NS}$
with the lattice expression provides a model value for $a_4$. The numerical result is $a_4^{(\sigma)}= -3.4\,\hbox{GeV}^{-3}$
while the value found in the expansion is only   $- 0.5\,\hbox{GeV}^{-3}$. 
\begin{figure}
\noindent
 \begin{minipage}[b]{.45\linewidth}
    \centering\epsfig{figure=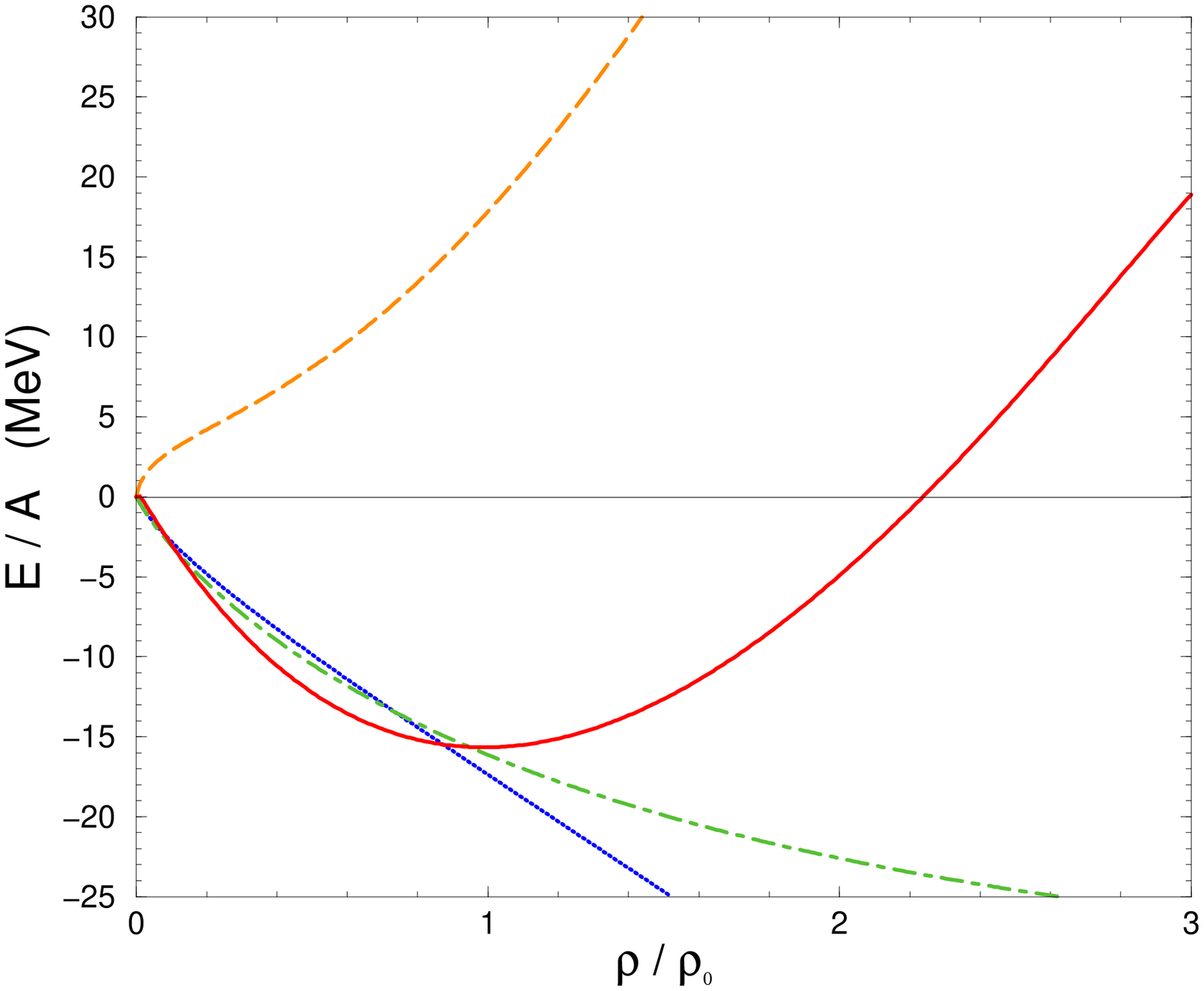,width=\linewidth,angle=270}
  \end{minipage}\hfill
  \begin{minipage}[b]{.45\linewidth}. 
  \centering\epsfig{figure=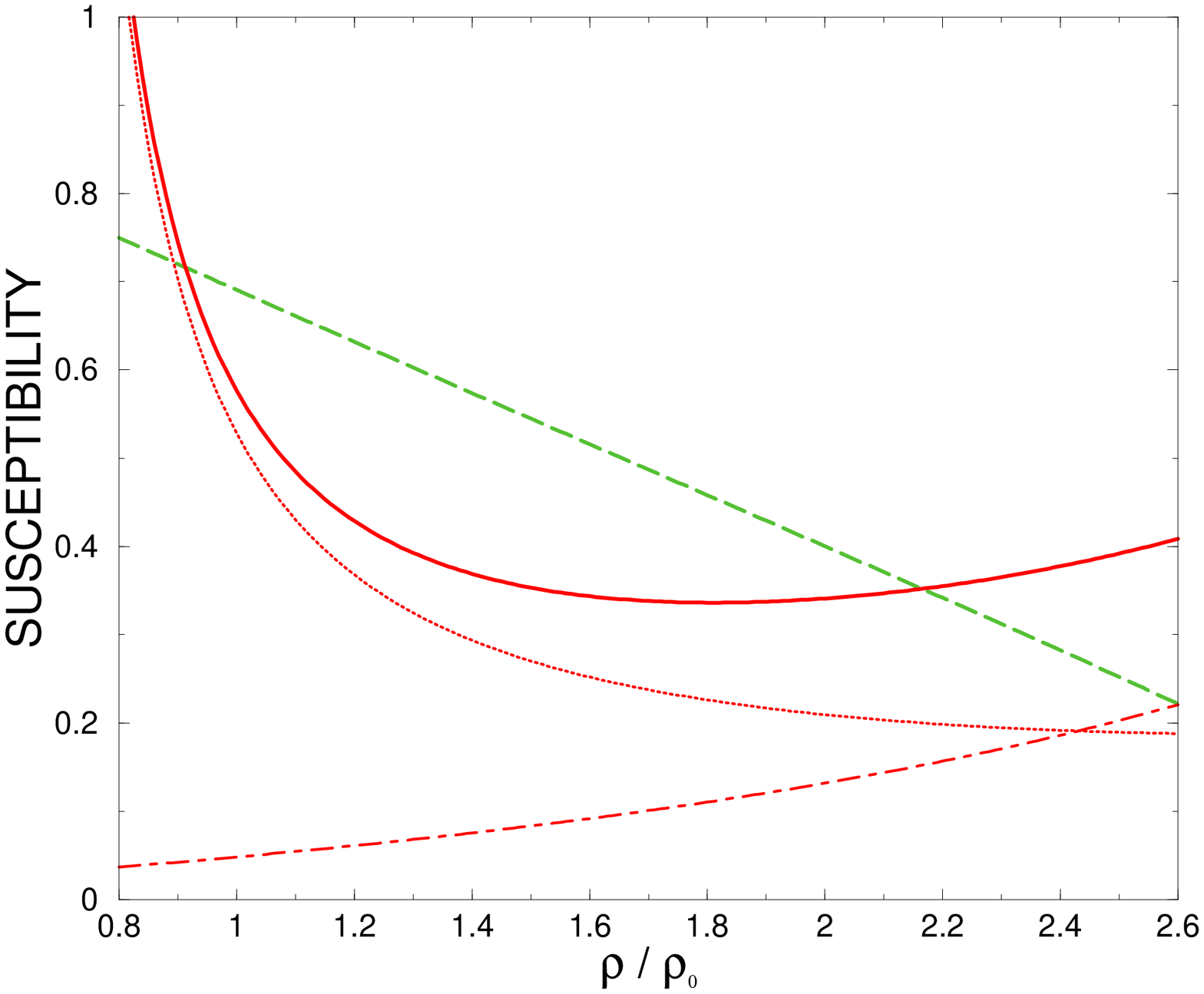,width=\linewidth,angle=270}
  \end{minipage} 
\caption{\it On the left panel the  binding energy of nuclear matter with $g_\omega=8.0$, $m_\sigma=850\, MeV$  and $C=0.985$ is shown.
The full line corresponds to the full result, the dotted line represents the binding energy
without the Fock and  correlation energies and the dot-dahed line corresponds to 
the contribution of the Fock terms. The decreasing dotted line (always negative) represents the
correlation energy. The right panel shows the density evolution of the QCD susceptibilities (normalized to the
 vacuum value of the pseudoscalar one).  
Dashed curve: pseudoscalar susceptibility. Full curve: scalar susceptibility. Dotted curve:
nuclear contribution to the scalar susceptibility. Dot-dashed curve: pion loop contribution to the
scalar susceptibility.  } 
\end{figure}
\begin{figure}
\noindent
 \begin{minipage}[b]{.40\linewidth}
    \centering\epsfig{figure=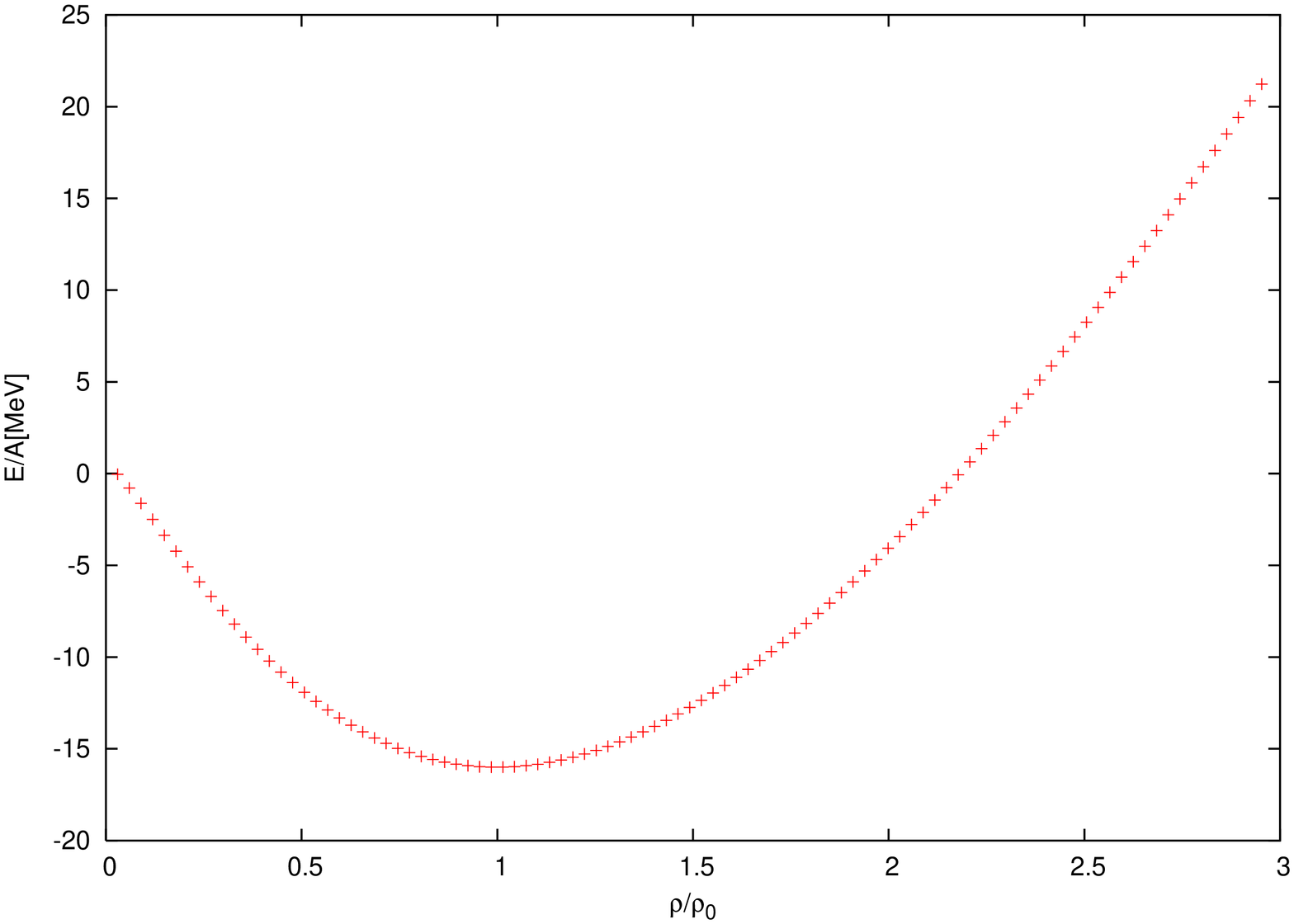,width=\linewidth,angle=270}
  \end{minipage}\hfill
  \begin{minipage}[b]{.40\linewidth}. 
  \centering\epsfig{figure=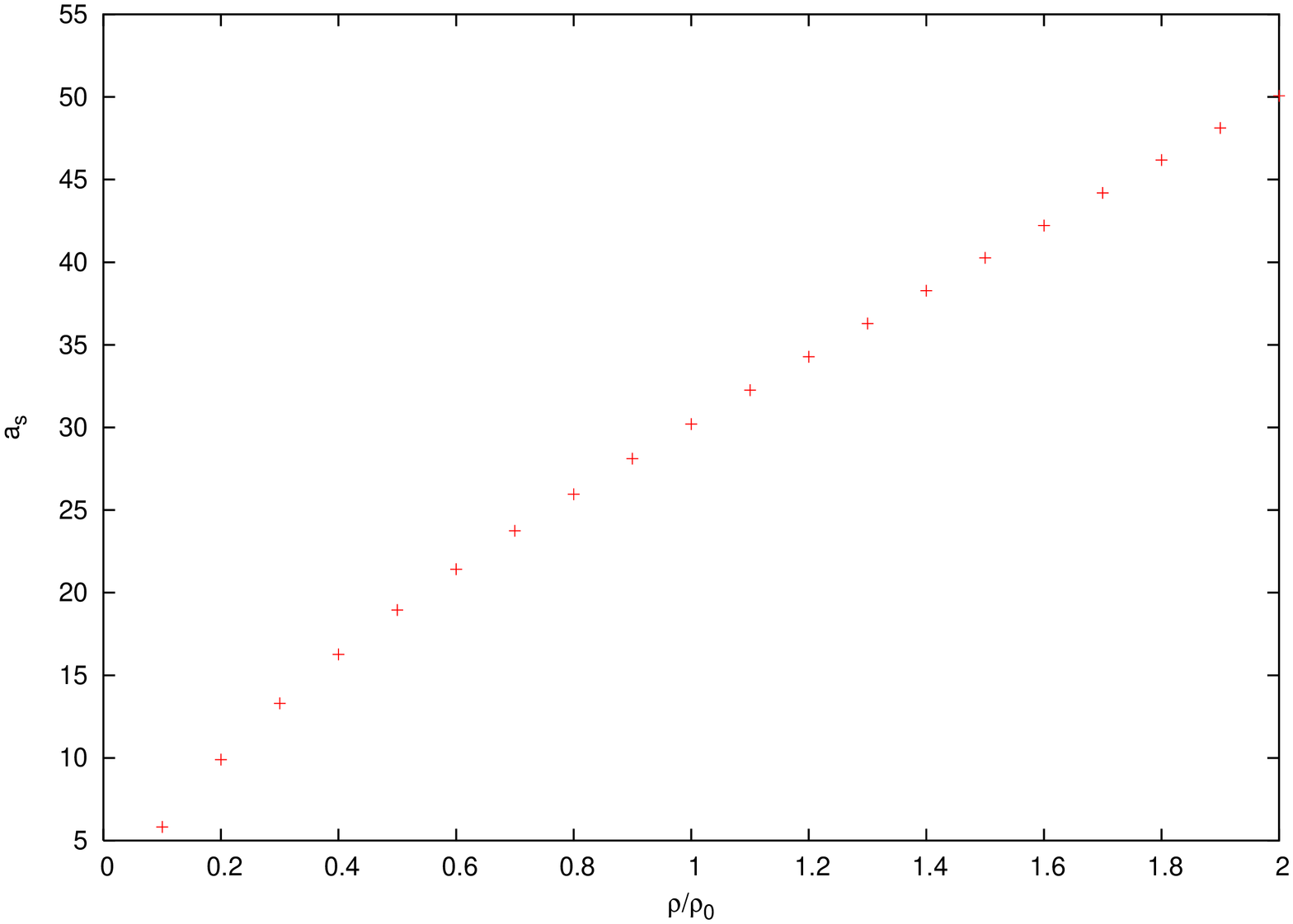,width=\linewidth,angle=270}
  \end{minipage}
\caption{\it Results of the RHF calculation with VDM parameters for the rho meson. Left panel: binding energy of symmetric nuclear matter. Right panel: density dependence of the asymmetry energy parameter. } 
\end{figure}

\bigskip\noindent
{\it Nucleon structure effects and  confinement mechanism}. The net conclusion of the above discussion is that the model as such fails to pass the QCD test. 
In fact this is to be expected and even gratifying because it also fails the nuclear physics test. We will see that these two important failures may have a common origin. Indeed  an important effect is  missing, namely  the  scalar response of the nucleon,  
$ \kappa_{NS}=\partial^2 M_N/\partial s^2$, to the scalar 
nuclear field, which is the basis of the quark-meson coupling model, (QMC), introduced in ref. \cite{G88}. The physical reason is very easy to understand: the nucleons are quite large composite systems of quarks and gluons and they should respond to the nuclear environment, {\it i.e.},
to the background nuclear scalar  fields. This response originates from the quark wave function modification  in the nuclear field 
and will oviously depend on the confinement mechanism. This confinement effect is expected to generate a positive scalar response 
$\kappa_{NS}$, {\it i.e.,} it opposes an increase of the scalar field, a feature   confirmed by the lattice analysis (see below).
This polarization of the nucleon is accounted for by the phenomenological introduction
of the scalar nucleon response, $\kappa_{NS}$, in  the nucleon mass evolution as follows~:
\begin{equation}
	M_N(s)=M_N\,+\,g_S\,s\,+\,\frac{1}{2}\,\kappa_{NS}\,s^2.
\end{equation}
This constitutes the only change in the expression of the energy density (eq. \ref{HARTREE}) but this has numerous consequences. The effective scalar coupling constant drops with increasing density but  the sigma mass gets stabilized~:
\begin{equation}
g^*_S(\bar s)={\partial M^*_N\over\partial\bar s}={M_N\over f_{\pi}}\,+\,\kappa_{NS}\, \bar s,\quad
m^{*2}_\sigma =\frac{\partial^2 \varepsilon}{\partial\bar s^2}\simeq
m^{2}_{\sigma}- ( {3\,g_S \over f_{\pi}} -  \, \kappa_{NS})\,\rho_S .
\end{equation}
The non-pionic  contribution to the nucleon susceptibility is modified, as well~:
\begin{equation}
 \chi_{NS}^{(\sigma)}= -2{\langle\bar q q\rangle_{vac}^2 \over f^2_{\pi} }\,
 \left({1\over m^{*2}_\sigma}\,-\,{1\over m^{2}_\sigma}\right)\,{1\over\rho}=
-2{\langle\bar q q\rangle_{vac}^2 \over f^2_{\pi} }\,{1\over m^{4}_\sigma}\,
\left({3\,g_S\over f_\pi}\,-\,\kappa_{NS}\right)\,. 
\end{equation}
We see that the effect of confinement ($\kappa_{NS}$) is to compensate  the pure scalar term. Again comparing with the lattice expression one gets 
a model value for the $a_4$ parameter~:
\begin{equation}
a_4= -{ a_2^2 \over 2\,M}( 3\,-\,2\,C).
\end{equation}
where $C$ is the dimensionless parameter $ C= \left({f^2_{\pi}/  2\,M} \right) \kappa_{NS} $. Numerically 
$a_4=- 0.5 \,GeV^{-3}$ gives $ C =+1.25$, implying a large cancellation.  As discussed below we will see that such a significant scalar response will generate other repulsive forces which restore the saturation mechanism. At this point it is important to come again to the underlying physical picture
implying that the nucleon mass originates both from the coupling to condensate and from confinement. In the original formulation of the quark coupling model, nuclear matter is represented as a collection of (MIT) bags seen as bubbles of perturbative vacuum in which quarks are confined. Thus 
in such a picture the mesons do not move and do not couple to quarks  as in the true non perturbative  QCD vacuum. Consequently the bag picture is at best an effective realisation of confinement which must not to be taken too literally. Indeed, QCD lattice simulations strongly suggest that a more realistic picture is closer to a $Y$ shaped color string (confinement aspect) attached to quarks. Outside this relatively thin string one has the ordinary non perturbative QCD vacuum possesing a chiral condensate from which the quarks get their constituant mass.

\section{Results for nuclear matter and discussion}

{\it Stability and chiral properties of nuclear matter}. In our first work devoted to the stability of nuclear matter \cite{CE05} we limited ourselves to the Hartree approximation (eq. \ref{HARTREE}). The free sigma mass and the $C$ parameter characterizing the nucleonic scalar response
were taken as adjustable parameters. The fact that the best fit came with values very close to what we deduced afterwards from lattice is certainly satisfactory. One problem was the  too large incompressibility $K$ but is  is nevertheless  possible to improve our model description  by adding an extra $s^3$ term in the expression of the effective nucleon mass in such a way that the nucleonic response vanishes at full chiral restoration ($\bar s=-f_\pi$). 
In this case the set of  parameters : $g_\omega=6.8$,  $m_\sigma=750$ MeV and $C=1$, leads to correct saturation properties, with an incompressibility value $K=270$ MeV. Another remarkable finding, according to the previous discussion,  was the  stability of the sigma 
mass which dropped by only $100$ MeV at $2.5 \rho_0$. Ignoring the repusive tadpole associated with the nucleonic response would have given an almost vanishing sigma mass at this density.

The next step \cite{CE07} has been to introduce pion loops which are necessary if we wish to address the question of the chiral properties of nuclear matter (namely the in-medium quark condensate and chiral susceptibilities) in a way which is consistent with nuclear matter stability since 
pion loops  affect  the energy. They  do not enter at 
the Hartree mean field level but  contribute through the Fock term and through the correlation energy. This energy can be  calculated within a RPA ring approximation according to~:
\begin{eqnarray}
E_{loop}=E\,-\,E_0&\equiv& V\,\varepsilon_{loop}={3}\,V\,\int_{-\infty}^{+\infty} 
{i\,d\omega\over 2\pi)}\int{d{\bf q}\over 
(2\pi)^3}\,\int_0^1{d\lambda\over\lambda}\nonumber\\
& &\left( \big[V_L(\omega, {\bf q})\,\Pi_L(\omega, {\bf q}; \lambda)\big]\,+\,2\,
\big[V_T(\omega, {\bf q})\,\Pi_T(\omega, {\bf q}; \lambda)\big]\right)~.\label{ELOOP}
\end{eqnarray}
$\Pi_{L}(\omega, {\bf q})$ is  the full (RPA) longitudinal spin-isospin polarization propagator in the pionic channel. This calculation
actually includes iterated pion exchange and also the part of the NN potential from the two-pion exchange 
with $\Delta$ excitation.
$V_L(\omega, {\bf q})$ is nothing but the (non static) pion exchange potential corrected by short-range components  
{\it via} the Landau-Migdal parameters, $g'_{NN},\, g'_{N\Delta},\, g'_{\Delta\Delta}$. We have taken their values from  
a systematic survey of the data on spin-isospin physics  by Ichimura {\it et al.} \cite{ISW06}~: $g'_{NN}=0.7,\,  g'_{N\Delta }=0.3,\,  
g'_{\Delta \Delta }=0.5 $. 
Beside pion exchange  we have also introduced the transverse channel ($V_T$), dominated by $\rho $ exchange together with the 
short-range component.  All the other needed ingredients are taken from the pion-nucleon and pion-nucleus phenomenolgy.  The scalar 
coupling constant is the one of the sigma model $g_S= M_N/f_{\pi} =10 $ and for the sigma mass we have followed the lattice 
indications, allowing a small readjustment around the lattice value, which is $m_{\sigma}=800\, MeV$. We have found a better 
fit with $m_{\sigma }=850\,MeV$ which corresponds to  $\sigma^{non-pion} = 26\,MeV$. The omega mass is known and the 
$\omega NN$ coupling constant is the only really free parameter. For the nucleon scalar response we have followed the indications of the 
lattice data but not strictly in view of the uncertainties attached to the higher derivatives. The value which fits 
the saturation properties is found to be $C\simeq 1$, not far from the lattice value $C= 1.25$. With these inputs 
we have obtained a satisfactory description of the nuclear binding.  The binding energy per particle is shown  in 
fig. 2, with its different components. Taking the successive derivatives of the equation of state (the grand potential $\omega=\varepsilon-\mu\rho$) with respect to $m^2_\pi$, it is possible to obtain the pseudo-scalar susceptibility (directly related to the quark condensate) and the scalar susceptibilities of nuclear matter \cite{CEG03}. Since they are associated with chiral partners (pion and sigma), they should be identical at chiral restoration. The numerical calculation \cite{CE07} on the right panel of fig. 2 indeed displays a convergence effect.

\bigskip\noindent
{\it Relativistic Hartree-Fock (RHF)}.
The description of asymmetric nuclear matter certainly necessitates the full rho exchange already at the Hartree level through its time component. We thus include it using the following interacting lagrangian~:
 \begin{equation}
	{\cal L}_\rho =- g_\rho\,\rho_{a\mu}\,\Psi\bar\gamma^\mu \tau_a\Psi
\,-\,g_\rho\frac{\kappa_\rho}{2\,M_N}\,\partial_\nu \rho_{a\mu}\,\Psi\bar\sigma^{\mu\nu}_{}\tau_a\Psi
\,+\,\frac{1}{2}\,m^2_\rho\,\rho_{a\mu}\rho^{\mu}_{a}
\,-\,\frac{1}{4} \,G_a^{\mu\nu}G_{a\mu\nu}
\end{equation}
One specific comment is in order for the tensor coupling of vector mesons. The pure vector dominance  picture (VDM)
implies the identification of $\kappa_{\rho}$ with the anomalous part of the isovector magnetic moment of the nucleon, {\it i.e.}, $\kappa_{\rho}=3.7$. However pion-nucleon scattering data suggest $\kappa_{\rho}=6.6$ (strong rho scenario).
At the Hartree level we still have  our previous picture. Symmetric and asymmetric nuclear matter are  seen as an assembly of nucleons, {\it i.e.,} of Y shaped color strings with massive constituant quarks at the end getting their mass from the chiral condensate.
This nucleons move in self-consistent scalar ($\bar s$) and vector background fields 
($\omega^0, \rho^0_3$).
However to get a quantitative  description, all the Fock terms, including not only pion and rho exchange but also sigma and omega exchange, are needed. These exchange terms are mediated by the propagation of the fluctuations of the previous meson fields. In particular it is possible to show that the scalar fluctuation, {\it i.e.}, the fluctuation of the chiral condensate, propagates with an in-medium modified sigma mass $m^{*2}_{\sigma}=V''(\bar s)\,+\,\kappa_{NS}\,\rho_S$.  The results of preliminary  RHF calculation \cite{MC07} with rho coupling given by pure VDM ($g_\rho=2.65$, $K_\rho=3.7$) are shown on fig. 3 and reproduce the saturation properties. $g_\omega=6.35$ is not very far from the quark model value $g_\omega=3\,g_\rho$ and $C=1.75$ is larger than the lattice value. The resulting incompressibility $K=242$ MeV and asymmetry energy parameter
$a_S=30.2$ MeV are obviously satisfactory. It is also possible to reproduce saturation properties in the strong rho scenario ($K_\rho=6.6$) but the asymmetry energy becomes too large~: $a_S=45$ MeV. These results are certainly very encouraging but the ultimate test of this chiral theory will be the reproduction of finite nuclei (in particular the spin-orbit splitting) and the consequences for neutron star properties.


\end{document}